\documentstyle[prd,aps,floats]{revtex}

\begin{document}
\preprint{SU-ITP-99/45\\ hep-ph/9910499\\ October 1999}
\draft

\input epsf
\renewcommand{\topfraction}{0.99}
\renewcommand{\bottomfraction}{0.99}

\def\be{\begin{equation}}
\def\ee{\end{equation}}
\def\ba{\begin{eqnarray}}
\def\ea{\end{eqnarray}}
\def\bq{\begin{quote}}
\def\eq{\end{quote}}
\def\PL{{ \it Phys. Lett.} }
\def\PRL{{\it Phys. Rev. Lett.} }
\def\NP{{\it Nucl. Phys.} }
\def\PR{{\it Phys. Rev.} }
\def\MPL{{\it Mod. Phys. Lett.} }
\def\IJMP{{\it Int. J. Mod .Phys.} }
\newcommand{\labell}[1]{\label{#1}\qquad_{#1}} 
\newcommand{\labels}[1]{\vskip-2ex$_{#1}$\label{#1}} 

\twocolumn[\hsize\textwidth\columnwidth\hsize\csname
@twocolumnfalse\endcsname

\title{Dynamics and perturbations in assisted chaotic inflation}
\author{Nemanja Kaloper}
\address{Department of Physics, Stanford University, Stanford CA 94305, USA}
\author{Andrew R.~Liddle}
\address{Astrophysics Group, The Blackett Laboratory, Imperial College,
London SW7 2BZ, UK\\
Isaac Newton Institute, University of Cambridge, Cambridge CB3 0EH, UK\\
(Address from 1st January 2000: Astronomy Centre,
University of Sussex, Brighton BN1 9QJ, UK)}
\date{\today}

\maketitle
\begin{abstract}
On compactification from higher dimensions, a single free massive
scalar field gives rise to a set of effective four-dimensional scalar
fields, each with a different mass.
These can cooperate to drive a
period of inflation known as assisted inflation. We analyze the
dynamics of the simplest implementation of this idea, paying
particular attention to the decoupling of fields from the slow-roll
regime as inflation proceeds. Unlike normal models of inflation, the dynamics
does not become independent of the initial conditions at late times. In
particular, we estimate the density perturbations
obtained, which retain a memory of the initial conditions even though a
homogeneous,
spatially-flat Universe is generated.
\end{abstract}
\pacs{PACS: 98.80.Cq \hfill  hep-ph/9910499, SU-ITP-99/45}

\vskip1cm]

\section{Introduction}

It was recently pointed out by Liddle et al.~\cite{lms} that, under
certain circumstances, it is possible for a set of scalar fields to
act cooperatively to drive a period of cosmological inflation, even if
none of the individual fields are capable of so doing.  Such behaviour
is known as {\em assisted inflation}, and in Ref.~\cite{lms}
investigation was made of the case of multiple fields moving in
exponential potentials with no interactions between them.  For
uncoupled fields this behaviour is generic; the physics is simply that
the fields feel the downward force from their own potential gradient,
but the collective friction from the whole set of fields through the
expansion rate $H$.  It was however perhaps unexpected that assisted
inflation solutions would be late-time attractors, confounding the expectation
that only the field with
the shallowest potential would dominate at late times.  The dynamics
of assisted inflation have subsequently been investigated by several
authors \cite{mw,pko1,cmn,gl}; one important additional point is that
direct interactions between the fields tend to inhibit assisted
behaviour.

A particularly simple implementation of assisted dynamics arises in
higher-dimensional theories with large compact internal spaces, as pointed out
by Kanti and Olive \cite{pko1,pko2}.  There have been many studies of
higher-dimensional
theories with compact internal spaces with size larger than the fundamental
scale, where the size could vary in the wide range between
a millimeter \cite{savas}, over ${\rm TeV}^{-1}$ \cite{anto}, up to only a few
orders of magnitude larger than the $4D$ Planck length \cite{hw,ovrut}.
Some aspects of universe as a domain wall have been considered
before \cite{domwall}. Most recently, a possibility that the extra
dimensions may even be in some sense infinite has been
considered \cite{rs}-\cite{inters}.
Cosmology in these theories may be very
different from the conventional $4D$ theory
\cite{earlyco,ahdkmr,moreco}.
A single fundamental scalar living in the bulk of
the higher-dimensional theory can give rise, upon compactification, to a set of
effective scalar fields corresponding to the Kaluza--Klein modes.  Provided the
extra dimensions are sufficiently large, this yields a large number of scalar
fields with similar potentials, which can support assisted inflation.  The main
advantage of this type of inflation is that individual fields need never exceed
the
fundamental Planck scale, which may allow supergravity corrections to remain
consistently small.  Another possible benefit of this scheme is that provided
the number of fields is sufficiently great, the fundamental self-couplings of
the fields may be much greater than the usual requirement that they be of order
$10^{-12}$ or less.

In this paper, we analyze the dynamics of a simple implementation of this idea,
where the scalar fields are uncoupled but have different masses depending on
the
Kaluza--Klein winding \cite{pko2}.
As inflation proceeds, the energy scale decreases, which leads to a reduction
in
the number of scalar fields with mass less than the Hubble scale.  Only such
fields can slow-roll.  Therefore, the number of fields participating in the
assisted behaviour decreases as inflation proceeds.  We will also study the
effect of this on the density perturbation spectrum.  Our results will reveal a
very interesting property of
assisted chaotic dynamics.  In contrast to the common
inflationary models, assisted chaotic models permit some information about the
initial conditions to be retained in the form of a soft dependence of the
spectral index, and other post-inflationary predictions, on the number of
fields
which contributed to inflation.

\section{The model}

Following Kanti and Olive \cite{pko2}, we consider a bulk theory of gravity in
five dimensions, with a
minimally-coupled massive scalar field $\Psi$.  The action is
\be
S_5 = \int d^5x \sqrt{g_5} \left[ \frac{R_5}{2\kappa^2_5}-\frac12 (\nabla
\Psi)^2 - \frac{1}{2} m^2 \Psi^2 \right] \,.
\label{act5}
\ee
The scalar field $\Psi$ is some weakly coupled light bulk field,
and we assume that its self-interactions are negligible compared
to the mass term. Such fields could arise in
some bulk supergravity theories.
After stabilization of the compactified dimension, the $5D$ metric
can be split as
\be
ds^2 = g_{\mu\nu} dx^\mu dx^\nu + L^2 d\theta^2 \,,
\label{metric5}
\ee
where $L$ is the constant size of the fifth dimension. The
relationship between the $5D$ and $4D$ parameters is easy to compute:
the $4D$ Planck mass, defined by $m^2_{{\rm Pl}}= 1/G$, is given in
terms of the $5D$ one as $m^2_{{\rm Pl}} = M^3 L$, where $M$,
is the $5D$ unification scale (the string scale). The wave function
renormalization for the scalar field and its Kaluza--Klein siblings is
$\Phi_j = \sqrt{L} \Psi_j $, where $\Phi_j$ is the projection of the
$j^{th}$ Kaluza--Klein state with mass
\begin{equation}
m^2_j = m^2 + \frac{j^2}{L^2} \,.
\label{KKmass}
\end{equation}
Of the infinite tower of the Kaluza--Klein states, only those which
admit a description in the field theory limit should be included here.
They are those states which are lighter than the fundamental scale of the
theory $M$. The states which are heavier than $M$ can be
described consistently only in the full quantum gravity limit, and
hence are left out. Then, from the mass formula Eq.~(\ref{KKmass}),
the total number of the light Kaluza--Klein states is $N_{{\rm max}} = ML =
m^2_{{\rm Pl}}/M^2$. Note that consistency requires $m \le M$, otherwise none
of
the
fields should be kept in the field theory limit.
Therefore the $4D$
reduced action is
\be
S_{4,{\rm eff}} = \int d^4x \sqrt{g} \left\{
\frac{m_{{\rm Pl}}^2 \, R}{16\pi} - \sum_{i\ge 0} \left[ \frac12(\nabla
\Phi_i)^2 +
\frac{1}{2}m^2_i \Phi_i^2 \right] \right\}.
\label{act4}
\ee
In this model the $4D$ scalar fields are coupled only through gravity, whereas
if the original $\Psi$ field were self-coupled the $\Phi_j$ fields would be
interacting.

Before exhibiting the calculation, let us overview the picture of
the evolution. This more or less follows that outlined by Kanti and Olive
\cite{pko2}, though they concentrated mainly on the lightest fields, whose
masses were taken to be nearly identical.
Assuming compactification has occurred, then, as in the chaotic
inflation paradigm \cite{andc,andh,book}, at early times the fields may range
over a wide variety of
values in different regions of space.  In the regime where the fields
do not exceed the $4D$ Planck scale, supergravity corrections are small.
In the original assisted picture using exponential potentials \cite{lms}, the
assisted inflation solution was the unique late-time attractor, with all fields
eventually participating. In this model the situation is rather different.
First, there is no late-time inflationary solution as eventually all fields
will
settle in the minima. Second, there is no attractor behaviour; as we will see
the fields tend to diverge from one another. Nevertheless, there is a form of
a transient assisted behaviour,
simply because there are many fields, and in particular the
light ones are in a slow-roll regime where they evolve slowly and feel the
collective friction of all the fields.

An important feature of this model is that the fields do not all have the same
potential; they receive a
contribution to their mass from the Kaluza--Klein winding $j$, and barring
severe fine tuning this mass difference will lead to significant
effects during evolution. This is because for each mode the
slow-roll regime can only be achieved if the mass of the mode $m_j$
is less than the Hubble parameter; otherwise the field will instead
evolve quickly to its minimum, with the energy density redshifting
as $1/a^3$ during the oscillations.  As assisted inflation proceeds, the
Hubble scale is decreasing, and so lighter and lighter fields enter
the regime $m > H$ and end their assisted behaviour.  Consequently
fewer and fewer fields are involved as time goes by.

Although there is no formal
attractor behaviour, nevertheless the collective effect of the fields may give
desirable advantages over single-field chaotic inflation.  The most prominent
of
these is that, due to the collective friction, inflation becomes possible when
all the fields have values less than the Planck mass.  This means that
supergravity corrections to the potential are much less likely to destroy
inflation than in the single-field case, where inflation is only possible for
$\phi \gtrsim m_{{\rm Pl}}$.  In the assisted variant, due to the collective
contribution to the friction, we will find that inflation is possible down to
field values $\Phi \sim m_{{\rm Pl}}/\sqrt{mL}$, where $m$ is the mass of the
zero
mode and $L$ the size of the extra dimensions.  As long as $mL \gg 1$, this is
well below the $4D$ Planck scale; slow-roll can occur for
sub-Planckian values of fields, and its duration is prolonged.  It is
interesting to note that because $L = m_{{\rm Pl}}^2/M^3$ \cite{savas},
inflation ends
at $\Phi \sim \sqrt{M/m} \, M$.  Thus for a fixed fundamental scale $M$, since
$m<M$, it is generally higher than $M$, but it lies lower the larger $m$ is.
So
even for relatively heavy fields, if they start out at $\Phi \sim m_{{\rm
Pl}}$,
there
may be many $e$-foldings of inflation provided $M < m_{{\rm Pl}}$.

\section{Inflationary dynamics}

We now consider the dynamics in detail.
Assuming a spatially-flat homogeneous cosmology, the equations of
motion for the assisted mass-driven chaotic inflation are
\ba
&&3H^2 = \frac{4\pi}{m^2_{\rm Pl}} \sum_{j=0}^{N(t)} \left( \dot \Phi_j^2 + m^2_j
\Phi^2_j\right) \,, \label{eom1} \\
&& \ddot \Phi_j + 3H \dot \Phi_j + m^2_j \Phi_j = 0 \,,
\label{eoms}
\ea
Although in principle the sum in the Friedmann equation goes over all
the fields up to $j = N_{{\rm max}}$, following the above discussion the only
fields which will
contribute in practice are those which are in the slow-roll regime. Therefore
we
can take the sum to only include those Kaluza--Klein states
whose mass is smaller than the Hubble parameter. We indicate this
number by $N(t)$, and it is time-dependent because $H$ is.

We are assuming initial conditions where all the fields are displaced from
their
minima. We will experiment with different choices.
While one could adopt the weak condition that the total energy
density is below the Planck scale, we prefer to take the more stringent
condition that supergravity corrections to the scalar field evolution can be
neglected. For the five-dimensional modes, this condition is
$\Psi_j(t_{{\rm initial}}) \lesssim M^{3/2}$, and after projecting to effective
$4D$ degrees of freedom this translates to $\Phi_j(t_{{\rm initial}}) \lesssim
m_{{\rm Pl}}$.  With a single field, this does not leave sufficient space for a
prolonged slow-roll regime, but with multiple fields slow-roll can continue
until $\Phi_j \ll m_{{\rm Pl}}$ \cite{pko2}.

\begin{figure}[t]
\centering
\leavevmode\epsfysize=8.5cm \epsfbox{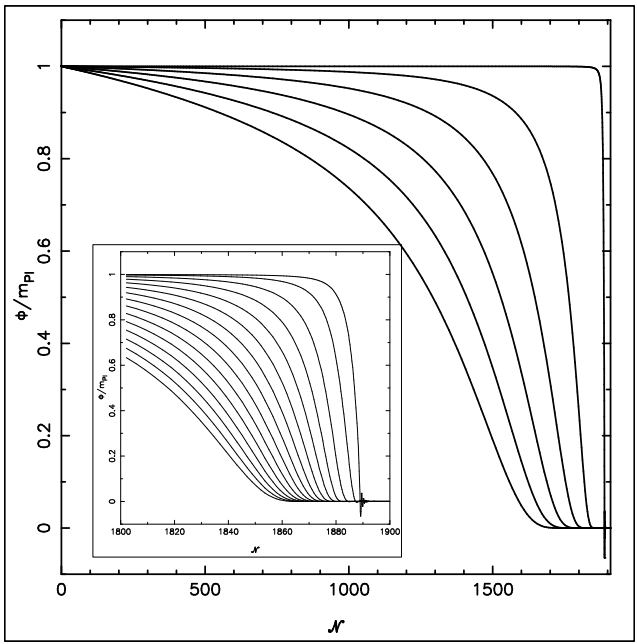}\\
\vspace*{5pt}
\caption[assfig1]{\label{f:assfig1} This shows the result of a 300 field
simulation, with $m = 10^{-4} \, m_{{\rm Pl}}$ and $L = 5000/m_{{\rm Pl}}$. The
initial condition for each field was
$\Phi_j = m_{{\rm Pl}}$, and evolution is shown as a function of ${\cal N}
\equiv \ln a$. The main panel shows the fields with $j = 0$, 25, 50, 75, 100 and
125, with the more massive fields decoupling first. The insert shows the lowest
15 fields at the end of inflation. Note that these fields are in the process of
decoupling during the last 50 $e$-foldings.}
\end{figure}

\subsection{Numerical solutions}

The system of equations described in Eqs.~(\ref{eom1}) and (\ref{eoms}) is
readily solved numerically, and we first describe the results of some
simulations to establish the general picture, before going on to describe some
analytical approximations which can be used.  For a given choice of the
parameters $m$ and $L$, and a choice of initial conditions used, one has to
decide how many scalar fields need to be evolved.  In general the more massive
ones will always swiftly become negligible, and so long as we restrict
ourselves to demanding an accurate description of only the
observably-accessible last 50 or so $e$-foldings of inflation, we need not
necessarily simulate the entire $N_{{\rm max}} = m_{{\rm Pl}}^2/M^2$ fields.

\begin{figure}[t]
\centering
\leavevmode\epsfysize=8.5cm \epsfbox{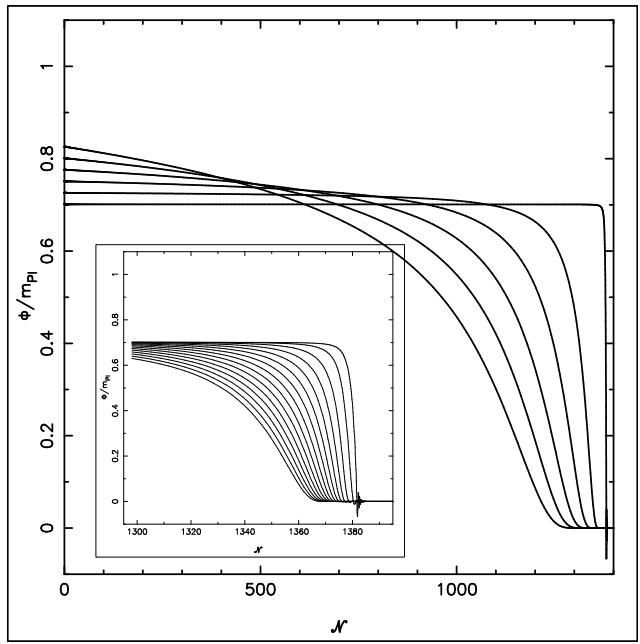}\\
\vspace*{5pt}
\caption[assfig2]{\label{f:assfig2} As Fig.~1, with the same model parameters
but now with different initial conditions for the fields.}
\end{figure}

Figs.~1 and 2 show two separate simulations of the same model ($m=10^{-4}
m_{{\rm Pl}}$ and $L = 5000/m_{{\rm Pl}}$) with different initial conditions.
With this $L$, $N_{{\rm max}} = (L\, m_{{\rm Pl}})^{2/3} \simeq 300$ so that
many
fields are included in the simulations. The fields are shown as a function of
the number of $e$-foldings of inflation ${\cal N} \equiv a$ (here given as the
number from the start of the simulation, rather than from the end of inflation).
In one simulation all fields start with the
same initial value, while in the other they have a spread of initial values.
The main panels show some fields up to quite large values of $j$, while the
insets show just the fifteen lightest fields. Note that the total number of
$e$-foldings is comfortably above the 70 or so required to solve the horizon
and flatness problems, and recall that it is only a few
$e$-foldings, centred around about 50 $e$-foldings from the end of inflation,
that can be directly probed by structure formation.

We see that the behaviour is indeed that outlined earlier. There is no evidence
of a late-time attractor, but the fields do take a substantial time to evolve
to the minimum and hence collectively drive inflation. In particular, we see
from the insets that at least the lightest 15 fields are still dynamically
relevant when structure-forming perturbations were imprinted. We also note that
after decoupling most fields simply asymptote into $\Phi_j = 0$; the other
fields contribute enough friction that the heavier ones remain overdamped right
to the minimum. Only the lightest few fields undergo oscillations when they
reach the minimum.

One vital result is to notice that, unlike usual inflation, the initial
conditions do not become irrelevant at late times, because in our region of the
Universe the fields have evolved from particular points in field space which
determines their time of decoupling. Studying Figs.~1 and 2, we see that the
evolution of the fields is not identical even during only the last 50
$e$-foldings. The late-time behaviour, and hence derived
quantities such as the density perturbation spectrum, will depend not only on
the underlying model parameters $m$ and $L$, but also on the particular initial
conditions for the fields pertaining to our region. This gives a significant
reduction in the usefulness of observations in directly constraining the
inflationary potential.

With these numerical results in mind, we now explore this
further using the slow-roll approximation.

\subsection{Analytical approximations}

At a given time $t$ all
the fields whose mass is smaller than $H$ are in the regime where the
collective friction dominates over the acceleration, and hence are
supporting inflation (perhaps with the exception of some which by chance
started
very near the origin). The mass of the heaviest contributing field is
given by Eq.~(\ref{KKmass}), with $j = N(t)$.  If we choose an initial
condition
that all the fields were approximately the same, one may suspect that by time
$t$ their values
may have become significantly different.  In fact, we can give an
accurate estimate of this spreading, and its effect on dynamics, as
follows. The slow-roll version of Eq.~(\ref{eoms}), obtained
by dropping the second derivative term, implies
\cite{pko2}
\be
\frac{\Phi_j(t)}{\Phi_j({\rm initial})} =
\left[ \frac{\Phi_0(t)}{\Phi_0({\rm initial})}\right]^{m^2_j/m^2_0}
\label{scaling}
\ee
where index $0$ refers to the zero mode, of mass $m$.
The initial values for all modes $\Phi_j({\rm initial})=\Phi_j(0)$
are taken to be of the order of $m_{{\rm Pl}}$, as we have discussed above.
However for the sake of generality we will retain them as
arbitrary set of input parameters, in order to study their
influence on the dynamics. Then, we can approximate
the $j^{th}$ mode by
\be
\Phi_j(t) = \Phi_j(0)\exp\left[-\left(1+ \frac{j^2}{m^2L^2}\right)
\sigma(t)\right]
\label{modesolutions}
\ee
where the field $\sigma$ is the logarithm of the zero mode,
$\sigma = -\ln[\Phi_0(t)/\Phi_0(0)]$ and the minimum of the potential
corresponds to $\sigma \rightarrow \infty$.

Then, assuming that there are many light fields in the slow-roll
regime so that we can replace the sum in Eq.~(\ref{eom1}) by an integral
$\int_0^{N(t)} dj$, and
ignoring their kinetic terms, after some simple algebra
we find that
\be
H^2 = \frac{4\pi}{3m_{{\rm Pl}}^2} \frac{m^3L}{\sqrt{2\sigma}} e^{-2\sigma}
\int_0^{\frac{\sqrt{2\sigma}N(t)}{mL}} dy f^2(y) \left(1+
\frac{y^2}{2\sigma}\right)
e^{-y^2} \,.
\label{approxfrid}
\ee
Here we have defined the function $f(y)$ by $f(y) =
\Phi_{mLy/\sqrt{2\sigma}}(0)$, since we can certainly view the initial
distribution of the Kaluza--Klein fields as a function of their mass.  In
practice, in most cases the upper limit of the integral can be taken to be of
order unity.  Indeed, at about $50$ $e$-foldings before the end of inflation,
the heaviest modes which are still in the slow-roll regime require $N(t) \sim
HL$.  On the other hand, by this time the field $\sigma$ will typically be of
order unity.  In fact, even if $\sqrt{2\sigma} \, N(t) \gg mL$, the integrand
falls rapidly to zero, as $\exp(-y^2)$, which effectively cuts off the
contributions to the integral to only those values for which $\sqrt{2\sigma} \,
N(t) \le mL$, as long as the function $f(y)$ grows slower than $\exp(y^2)$.
But
this is a very natural assumption in the assisted inflation context:  initially
all the fields which contribute to the collective attractor should have similar
values, and in fact the heavier fields should be lower along the potential
well.
Therefore the function $f(y)$ should be bounded by $f(y) \le f(0)$.  Hence
$f^2(y)$ in the integrand can be approximated by a polynomial, $f^2(y) \sim
\Phi^2_0(0) + p(y)$, which changes slowly compared to $\exp(-y^2)$.  So for as
long as inflation is proceeding with many fields still in the slow-roll regime,
we can safely replace the control parameter $\sqrt{2\sigma} \, N(t)/mL$ by a
number $\simeq 1$ in the upper limit of integration.

On the other hand, for a certain subspace of the phase space of the theory, the
quantity $\sqrt{2\sigma} \, N(t)/mL$ can be smaller than unity.  Typically this
happens for the choice of parameters in the Lagrangian where the zero mode is
heavy, and so most of its Kaluza--Klein siblings are decoupled.  Therefore
although generically assisted dynamics leads to many more $e$-foldings than the
minimum of $60$, for sufficiently large mass $m$, the total number of
$e$-foldings can be small.  In this case the appropriate approximation for
evaluating Eq.~(\ref{approxfrid}) is to take ${2\sigma}\, N(t)/mL \ll 1$, which
we label as short assisted inflation.  We will return to this case below.  Here
we must underscore that due to the strong nonlinear nature of the dynamics, we
need to treat the approximations in a floating manner.  To decide which
approximation is applicable at ${\cal N}$ $e$-foldings before the end of
inflation, we must check the value of the control parameter $\sqrt{2\sigma} \,
N(t)/mL$, and choose the relevant formulae for the spectra of perturbations as
it dictates.

\subsubsection{Long assisted inflation}

Armed with the above, in the case of long assisted inflation
we can estimate the integral by using the error
function ${\rm Erf}[x] = \frac{2}{\sqrt{\pi}} \int^{x}_0 dy \, e^{-y^2}$.
Since, as we said, the main contribution comes from the range of
values for which $x \le 1$, we can use $\int^1_0 dy e^{-y^2} \sim
\sqrt{2} c$, where $c$ is a number of order unity. The precise value
of $c$ is not of immediate consequence here,
and we will keep it as a free parameter for now.
This gives
\be
H^2 \simeq \frac{4 \pi c}{3} \frac{\Phi_0^2(0)}{m^2_{{\rm Pl}}}
\frac{m^3L}{\sqrt{\sigma}} e^{-2\sigma}.
\label{approxfridres}
\ee
Using the definition of $\sigma$, we can now rewrite this
equation in terms of the zero mode $\Phi_0(t)$, finding
\be
H \simeq \sqrt{\frac{4\pi }{3}}
\frac{(cmL)^{1/2}}{m_{{\rm Pl}}\ln^{1/4} \left[\Phi_0(0)/\Phi_0(t)\right]}
m \Phi_0(t).
\label{Hphi}
\ee
This equation permits us to replace the collection of fields
by the zero mode $\Phi_0$. Note the logarithmic dependence of
the Hubble parameter on the initial value of the zero mode field $\Phi_0$. This
takes into
account the decoupling of the heavy modes, which fall out of
the slow-roll regime as $\Phi_0(t)$ rolls towards the minimum.
Also note that because of the strong cutoff in the integration
in Eq.~(\ref{approxfrid}) effected by $\exp(-y^2)$, the dynamics
is sensitive only to the average of the initial
values of the Kaluza--Klein fields $\Phi_0(0)$, and not to its
dispersion. Hence, we can indeed safely assume that all the fields
were initially essentially the same.

Therefore to the lowest order we can rewrite the equations of motion,
using the slow-roll approximation, as
\ba
&&3H^2 = 4\pi \frac{cmL}{m^2_{{\rm Pl}}\ln^{1/2} \left[ \Phi_0(0)
/\Phi_0(t)\right]} m^2
\Phi^2_0(t)
\,, \label{sreom1} \\
&&3H \dot \Phi_0(t) + m^2 \Phi_0(t) = 0 \,.
\label{sreom2}
\ea
Since the field $\Phi_0$ is in slow-roll as long as $m \le H$, it
is now easy to see that inflation continues as long as
\be
\Phi_0 \ge \sqrt{\frac{3}{4\pi}} \frac{m_{{\rm Pl}}}{\sqrt{cmL}}
\ln^{1/4}\left(\sqrt{\frac{4\pi cmL}{3}}\frac{\Phi_0(0)}{m_{{\rm Pl}}}\right)
\,.
\label{phidurbound}
\ee
This equation is illustrative,
since now we indeed see, as we have mentioned above, that
in most cases during inflation
\be
\sqrt{2\sigma} \frac{N}{mL} \simeq \sqrt{2\sigma}
\frac{H}{m} \ge \sqrt{2}\ln^{1/2}(\Phi_0(0)/\Phi_0) \ge 1 \,,
\ee
except during the first few $e$-foldings immediately after the start.
One should bear in mind that the `first few' $e$-foldings
could in principle be enough to solve all the usual
cosmological problems. However, this will not be the
case for the most allowed values of the zero mode parameters.
Let us define
the quantity
\be
\alpha \equiv \frac{cmL}{\ln^{1/2}[\Phi_0(0)/\Phi_0(t)]} \,,
\label{alphadef}
\ee
which is a slowly-varying function of time through the time-dependence of
$\Phi_0$.
Now,
we can combine Eqs,~(\ref{sreom1}) and (\ref{sreom2}), and after some
straightforward algebra obtain
\be
\frac{d{\cal N}}{d\Phi_0} = 4\pi \alpha \, \frac{\Phi_0}{m_{{\rm
Pl}}^2} \,,
\label{Nphi}
\ee
where ${\cal N} =  \ln(a({\rm final})/a)$ is the number of $e$-foldings of
inflation which
occur after the field reaches $\Phi_0$ (not to be confused with $N(t)$, the
number of slow-rolling fields). It is straightforward to integrate this
equation: we find
\be
{\cal N} = (2 \pi)^{3/2} c\, mL \frac{\Phi^2_0(0)}{m^2_{{\rm Pl}}}
\left\{ 1- {\rm Erf}\left[\sqrt{2}
\ln^{1/2}\left(\frac{\Phi_0(0)}{\Phi_0}\right)\right]
\right\} \,.
\label{ne-foldings}
\ee
Since initially $\Phi_0(0) \sim m_{{\rm Pl}}$,
this equation shows that the total number of $e$-foldings
is ${\cal N}({\rm total}) \sim (2\pi)^{3/2} c\, m L $.
Further, it can be seen that at a time considerably before
the end of inflation, the number of $e$-foldings left
to leading order scales as ${\cal O}(\Phi^2_0)$. Indeed,
ignoring the variation of the denominator with $\Phi_0$ in Eq.~(\ref{Nphi}),
it follows that at ${\cal N} \gg 1$ $e$-foldings before the end of inflation
we can approximate Eq.~(\ref{ne-foldings}) with
\be
{\cal N} \simeq 2\pi\alpha \, \frac{\Phi^2_0}{m^2_{{\rm Pl}}} +
{\cal O}\left(\frac{\Phi^3_0}{m^3_{{\rm Pl}}}\right) \,.
\label{apprnefolds}
\ee
This formula is familiar from the usual chaotic inflation, except for
the factor $\alpha$ which comes
from the collective dynamics. In fact, this formula gives an
accurate approximation for the relationship between ${\cal N}$ and
$\Phi_0$ sufficiently far before the end of inflation, and we will use
it hereafter.

\subsubsection{Short assisted inflation}

Let us now return to the case of short assisted inflation.
As we will see later it is helpful to place
an upper bound on the mass of the zero mode $m$ which leads to
significant assisted behaviour.
In this instance, the integral Eq.~(\ref{approxfrid}) is better
approximated by
\be
H^2 =\frac{4\pi m^2}{3m_{{\rm Pl}}^2} N \Phi^2_0
\left(1+ \frac{N^2}{3m^2L^2}\right)
\,.
\label{approxfridsmall}
\ee
Since $N \sim HL$, using Eq.~(\ref{approxfridsmall}) we find
\be
N \simeq \frac{4\pi m^2L^2}{3 m^2_{{\rm Pl}}} \Phi^2_0 \left(1+
\frac{N^2}{3m^2L^2}\right)\,.
\ee
Now, although the second term on the RHS of this equation dominates,
we will approximate $N$ by the first term on the RHS, since
using the second
term in the subsequent framework would lead to an overestimation.
By using the first term,
the computation remains confined in the realm of perturbation theory,
where errors are controllable.
Thus $N \simeq 4\pi m^2 L^2 \, \Phi_0^2/3m^2_{{\rm Pl}}$.
Therefore,
\be
H^2 =\frac{16\pi^2 m^4 L^2}{9m_{{\rm Pl}}^4} \Phi^4_0
\left(1+ \frac{16\pi^2 m^2 L^2 \Phi^4_0}{9m^4_{{\rm Pl}}}\right)
\,.
\label{approxfridsmalla}
\ee
Next, we will approximate the Hubble parameter $H$ by
retaining the second term on the RHS of this equation, since
it clearly dominates over the first, while the perturbation
theory is still valid using it. Hence taking the square
root of Eq.~(\ref{approxfridsmalla}) we find
\be
H =\frac{16\pi^2 m^3 L^2}{9m_{{\rm Pl}}^4} \Phi^4_0
\,.
\label{approxfridsmallb}
\ee
Clearly since inflation lasts as long as $m \le H$, during it
$\Phi_0 \ge \sqrt{3/4\pi} \, m_{{\rm Pl}}/\sqrt{mL}$,
which is in good agreement with Eq.~(\ref{phidurbound}).
We note that the authors of Ref.~\cite{pko2}
treat assisted dynamics only in this regime where $\sqrt{\sigma} N/mL \ll 1$,
as is evident from their equations of motion. However, their
approximation for the Hubble parameter consists of retaining
only the linear term in Eq.~(\ref{approxfridsmall}).
The treatment here provides a more precise approximation.
Then using Eq.~(\ref{eoms}), we obtain
\be
\frac{d{\cal N}}{d\Phi_0} = \frac{256 \pi^4 m^4L^4}{27m^8_{{\rm Pl}}} \Phi^7_0
\,.
\label{smallphieq}
\ee
Its solution can be found immediately: it is
\be
{\cal N }= \frac{32 \pi^4 m^4L^4}{27m^8_{{\rm Pl}}} \Phi^8_0 \,.
\label{smallsol}
\ee
This equation replaces Eq.~(\ref{apprnefolds}) when $m$ is large,
or equivalently when the parameter $\alpha$ given in Eq.~(\ref{alphadef})
is small. We will consider this case in more detail in the next section.
Note that the approximations in this subsection give a good
description of the assisted dynamics for the times when
the control parameter $\sqrt{2\sigma} \, N(t)/mL$ is small,
initially or by subsequent dynamics.

The previous analysis shows that the
dissipation due to the decoupling of the massive
modes is the dominant correction to the dynamics of the
model. To estimate it, we need to establish how rapidly
the modes decouple in the course of evolution.
This can always be done by using the rule $N \sim HL$ and
the solutions presented so far.
In general, although it is clear
that the details are complicated, we
can nevertheless obtain a good approximation
for the decoupling rate as
\be
\frac{dN}{N} = \frac{d{\cal N}}{\gamma {\cal N}},
\label{decouprate}
\ee
where $\gamma$ measures the decoupling rate, and depends on the
parameters $m$ and $L$ and the initial conditions.  In
general it is a slowly-changing function, with the value within an
order of magnitude or two of unity. Note, that the usual single-field chaotic
inflation corresponds to the limit $\gamma \rightarrow
\infty$.  This is sufficient for our purposes here.

\section{Perturbations}

\subsection{Density perturbations}

Now we compute the density contrast, using the notation of
Ref.~\cite{LL}. We will do this approximately, by using the usual formula as
applied to the lightest scalar field. Unfortunately this ignores the effects of
perturbations in the other fields, but in the absence of exact analytical
solutions it is unclear how to include them. Ideally, one would follow the
approach of Malik and Wands \cite{mw}, in which a new set of fields is defined
such that the linear perturbations in all but one give no first-order
contribution to the perturbation in the total density, but this requires
knowing
the full solutions in advance. We hope to return to the question of a complete
computation of the adiabatic density contrast in a later work.

With the above caveats, the power spectrum ${\cal N}$ $e$-foldings before the
end of inflation is estimated as
\ba
\delta_{{\rm H}}(k) = \frac{1}{5\pi} \frac{H^2}{\dot{\Phi}_0}
& = & \sqrt{\frac{64\pi}{75} } \,
\frac{(cmL)^{3/2}}{\ln^{3/4}[\Phi_0(0)/\Phi_0]} \,
\frac{m \Phi_0^2}{m^3_{{\rm Pl}}} \nonumber \\
& = & \sqrt{\frac{16}{75\pi}} \, \sqrt{\alpha} \, \frac{m}{m_{{\rm Pl}}}
\, {\cal N} \,.
\label{contrastn}
\ea

As always with inflation, the overall amplitude can be adjusted to match the
value observed by COBE, $\delta_{{\rm H}}(k_{{\rm hor}}) \simeq 2 \times
10^{-5}$, by suitable choice of the mass parameter. Assuming that the present
Hubble radius equaled the Hubble radius 50 $e$-foldings from the end of
inflation, this requires
\be
\alpha_{50} = 2.3 \times 10^{-12} \, \frac{m_{{\rm Pl}}^2}{m^2} \,,
\label{alfifty}
\ee
where $\alpha_{50}$ is the value of $\alpha$ at the relevant epoch.
The model
has three input parameters, $m$, $L$,
and $\Phi_0(0)$, though the dependence on
the last is weak.

Another bound on the parameters can be obtained from
considering the initial energy density.
Here we will at first ignore the precise aspects of the
nonlinear assisted dynamics in order to estimate
the range of allowed values of the zero mode mass.
We will then reconsider some of the resulting inequalities
with more precision, since they will provide
the criteria for choosing the relevant approximation
for computing the spectral properties.
By the usual
arguments in theories with large extra dimensions \cite{savas},
the total energy density in $5D$ cannot exceed $M^5$.
After the extra dimensions are stabilized, this places an
upper bound on the projected energy density in $4D$:
$\rho \le M^5 L = M^2 m^2_{{\rm Pl}}$. On the one hand, the
initial energy density of the collective
inflaton is $\rho \sim m^3 L m^2_{{\rm
Pl}}$ by
Eq.~(\ref{approxfrid}). This
and the upper bound give us $m^3L \le M^2$. On the
other hand, by the definition of
$\alpha$ in Eq.~(\ref{alphadef}), ignoring the
logarithm in the
denominator, we can see that Eq.~(\ref{alfifty}) implies
$m^3L \sim 2.3 \times 10^{-12} m^2_{{\rm Pl}}$.
Hence combining the inflationary and
compactification constraints, we find
the lower bound on the fundamental scale $M$:
\be
M \ge 1.5 \times 10^{-6} m_{{\rm Pl}} \,.
\label{scalebound}
\ee
However, using Eq.~(\ref{alfifty}) again,
and noting that $m^3L = m^2_{{\rm Pl}} m^3/M^3$, we find that
$m = 1.32 \times 10^{-4} M$. Hence, combining this and
inequality (\ref{scalebound}), we obtain a lower bound
for the mass of the zero mode $m$:
\be
m \ge 2 \times 10^{-10} m_{{\rm Pl}}\,.
\label{massbound}
\ee
Therefore we see that the assisted models of inflation
could be a phenomenologically viable scenario of inflation only in
theories with a unification scale $\ge 10^{13} \, {\rm GeV}$, which is still
considerably
larger than the electroweak scale. Otherwise, assisted chaotic inflation
would not produce the density contrast in the COBE range.
This agrees with the conclusion that inflation after
stabilization of extra dimensions could give density
contrast as measured by COBE only if the unification
scale is high \cite{earlyco,ahdkmr}.

From this we find that the parameter $\alpha_{50}$ is
bounded from above: combining (\ref{alfifty}) and
(\ref{massbound}) we obtain
\be
\alpha_{50} \le 5.75 \times 10^{13}\,.
\label{alphabound}
\ee
Since $\ln[\Phi_0(0)/\Phi_0] \le \ln(2\pi\alpha/{\cal N})/2 \sim 15$,
the approximation made above in ignoring the logarithm
was justified. Finally from the fact that the total number of
$e$-foldings is ${\cal N}({\rm total}) \sim (2\pi)^{3/2} mL$, we see that
${\cal N}({\rm total}) \le 3 \times 10^{8}$, implying that
generically it can be quite large.

The most useful predictions that can then be made are of the shape of the
spectrum, which is independent of the normalization.  As compared to
single-field chaotic inflation, the density contrast in assisted chaotic
inflation has additional dependence on the inflaton via the factor $\alpha$,
which will cause a deviation of the shape of the spectrum.

We begin by calculating the spectral index $n$ of the spectrum. This is defined
by
\be
n-1 = \frac{d \ln \delta_{{\rm H}}^2(k)}{d \ln k} \,,
\ee
where in the slow-roll approximation $d\ln k \simeq - d{\cal N}$. The
interesting thing is that, unlike the usual case, this
scale-dependence receives two contributions. The first is the usual
one coming from the ${\cal N}$ term, and the second arises from the ${\cal
N}$-dependence of $\alpha$.

If initially we assume $\alpha$ is constant, we immediately find,
using Eq.~(\ref{apprnefolds}), the usual result for single-field quadratic
chaotic inflation, namely
\be
n = 1 - \frac{2}{{\cal N}}.
\label{tiltchao}
\ee
We can also compute the scale-dependence of the spectral index
\be
\frac{dn}{d\ln k} \simeq
- \frac{dn}{d{\cal N}} = - \frac{2}{{\cal N}^2} \,.
\label{spchao}
\ee
If we assume the present Hubble scale equaled the Hubble
scale $50$ $e$-foldings before the end of inflation, we have $n =
0.96$ and $dn/d\ln k = -8 \times 10^{-4}$. The Planck satellite is
capable of distinguishing the former from unity \cite{TEHC}, but not the latter
from zero \cite{CGL}.

However, we need to include the speed up of the inflaton field as
heavy Kaluza--Klein modes fall out of the slow-roll regime, through the
$\alpha$
term. From Eqs.~(\ref{apprnefolds}) and (\ref{contrastn}), we find
\be
n = 1 - \frac{2}{{\cal N}} \left[ 1
+ \frac{3}{4\ln[(2\pi \alpha \Phi^2_0(0))/({m^2_{{\rm Pl}}\cal N})]} \right]
\,.
\label{tiltassist}
\ee
Its scale-dependence is given by the equation
\ba
\frac{dn}{d\ln k} &=& -\frac{2}{{\cal N}^2} \Bigl[1
+ \frac{5}{4\ln[(2\pi \alpha \Phi^2_0(0))/({m^2_{{\rm Pl}}\cal N})]} \,
\nonumber \\
&&- \frac{3}{8\ln^2[(2\pi \alpha \Phi^2_0(0))/({m^2_{{\rm Pl}}\cal N})]} \Bigr]
\,.
\label{spassist}
\ea
In the limit $\alpha \rightarrow
\infty$, these expressions correctly reduce to Eqs.~(\ref{tiltchao}) and
(\ref{spchao}), as appropriate for single-field models.  However, for
a generic assisted model, we see that the predictions for the spectral
index and its scale dependence are different from the usual single-field
chaotic
inflation.
For example, if $\Phi_0(0) \sim m_{{\rm Pl}}$ and
the mass of the zero mode $m$ is of order of
$m \sim 10^{-7} m_{{\rm Pl}}$, the parameter $\alpha_{50}$ is,
using Eq.~(\ref{alfifty}),
$\alpha_{50} = 230$, and hence the spectral index and its gradient are
$n = 0.95$ and $dn/d\ln k = - 1.1 \times 10^{-3}$, respectively.
Clearly, these numbers are sensitive to the mass $m$: the smaller it
is, the more similar assisted chaotic inflation becomes to the usual
single-field driven chaotic inflation.

A closer look at Eqs.~(\ref{tiltassist}) and (\ref{spassist})
shows that the predictions depend on the initial condition $\Phi_0(0)$
through the parameter $\alpha$. Hence the spectral
index can vary significantly with the initial condition. The explicit
dependence of $n$ (and $dn/\ln k$) on the initial condition can be
computed from the above formulae. The important observation is that
since the initial value of the field $\Phi_0(0)$ appears only
through the logarithm, if $\Phi_0(0)$
varies through its full range of admissible values,
$ 10^{-4} m_{{\rm Pl}} \le \Phi_0(0) \le m_{{\rm Pl}}$, the
logarithm changes by a factor of $\sim 3$.
Hence, $n$ may vary in the range $0.94 \le n \le 0.96$,
at the $50$ $e$-foldings. In a manner of speaking, the assisted chaotic
inflation does not impart amnesia on the universe as efficiently
as single-field chaotic inflation, and some of the information
about the initial state of the universe much before the last $60$ $e$-foldings
is imprinted on the late epoch too.

We note that Eqs.~(\ref{tiltassist}) and (\ref{spassist}) suggest
that for the value of the mass $m$ where $\alpha_{50} = 25/\pi$ there is a
divergence. But this divergence is clearly completely spurious:
it merely signifies that at the large values of the mass $m$ the
approximations which led to Eqs.~(\ref{tiltassist}) and (\ref{spassist})
break down. Instead, there we need to resort to the approximation for
short assisted inflation, discussed at the end of the previous section.
Using Eqs.~(\ref{approxfridsmallb}) and (\ref{smallsol}),
it is then easy to derive the density contrast in this case:
\ba
\delta_H &=& \frac{4096 \pi^5 m^6L^6}{1215 m^{11}_{{\rm Pl}}} \frac{m}{m_{{\rm
Pl}}}
\Phi^{11}_0 \nonumber \\
&=& 1.5 \sqrt{mL} \frac{m}{m_{{\rm Pl}}} {\cal N}^{11/8}\,.
\label{contrastsmall}
\ea
Then, the COBE normalization condition gives
\be
mL = 3.7 \times 10^{-15} \frac{m^2_{{\rm Pl}}}{m^2}\,.
\label{cobenorm}
\ee
The spectral index is
\be
n = 1 - \frac{11}{4{\cal N}}\,,
\label{spindsmall}
\ee
and its gradient is
\be
\frac{dn}{d\ln k} = - \frac{11}{4{\cal N}^2}\,.
\label{gradsmall}
\ee
From the requirement that there is at least $60$ $e$-foldings
and Eqs.~(\ref{smallsol}) and (\ref{cobenorm}) we
can deduce the upper bound on the mass $m$. Clearly,
the maximal number of $e$-foldings will come from the largest
initial condition, $\Phi_0(0) \simeq m_{{\rm Pl}}$. Thus,
${\cal N} \simeq 50$ gives $m \sim 10^{-7} m_{{\rm Pl}}$, and the
spectral index and its gradient are $n = 0.945$ and
$dn/d\ln k = -1.1 \times 10^{-3}$. These numbers are in a very good agreement
with the corresponding numbers for the smallest attainable parameter
$\alpha_{50}$ discussed above. Therefore, since we limit
the initial condition of the field $\Phi_0(0)$ to be below the
$4D$ Planck scale $m_{{\rm Pl}}$, the largest mass $m$ which still
leads to sufficient inflation is $m \sim 10^{-7} m_{{\rm Pl}}$.
In this case the assistance effect produces a different
spectrum of perturbations from the usual single-field
chaotic inflation with quadratic potential.

We now turn to specifying the criteria for selecting
the relevant approximation on the basis of the values
of $m$. Since the control parameter is
\begin{equation}
{\cal C} = \sqrt{2\sigma}\frac{N}{mL} \sim \sqrt{2\sigma}\frac{H}{m} \,,
\end{equation}
using Eqs.~(\ref{Hphi}) and (\ref{apprnefolds}), we can rewrite it as
${\cal C} \sim \sqrt{4 \sigma {\cal N}/3}$.
Hence at ${\cal N} = 50$ $e$-foldings, this gives
${\cal C}_{50} \sim 8 \sqrt{\sigma_{50}}$, and
using the relationship between $\sigma_{50}$ and  $\alpha_{50}$,
and $M^3L = m^2_{Pl}$, we finally find
\be
{\cal C}_{50} \sim 3c \times 10^{11} \frac{m^3}{M^3}.
\ee
The parameter $c$ is never larger than $\sqrt{\pi/8} \sim 0.6264...$, and
while it can be smaller, we see that the control parameter
is really dominated by the ratio of the zero mode mass to the $5D$
Planck scale. This is a consequence of the
higher-dimensional origin of the effective inflaton field, as we have
discussed before. Therefore, we finally have a clear-cut criterion
for selecting one of the two approximate descriptions discussed above:
if ${\cal C}_{50} \gg 1$ we cannot ignore the nonlinearities
and should use the approximations for the long assisted inflation,
whereas for ${\cal C}_{50} < 1$ we can use the (improved) quasi-linear
approximations appropriate for the short assisted inflation.
Clearly, there is a transition region in between, where neither
of our approximations will be very accurate, and where the
complete treatment of the evolution necessitates a numerical
approach. However, this occurs
on only a small part of the phase space, while the analytical
approximations which we have developed
cover most of the admissible parameter space.

\subsection{Gravitational waves}

The gravitational waves produced during inflation are determined entirely by
the
evolution of the Hubble parameter. Following the notation of
Ref.~\cite{llkcba},
their amplitude is given by
\be
A_{{\rm G}} = \frac{2}{5\sqrt{\pi}} \, \frac{H}{m_{{\rm Pl}}} =
\sqrt{\frac{16}{75}} \, \frac{(cmL)^{1/2}}{\ln^{1/4}[\Phi_0(0)
/\Phi_0]} \, \frac{m \Phi_0}{m^2_{{\rm Pl}}} \,.
\label{gravwave}
\ee
As the gravitational wave production depends only on the expansion rate $a(t)$,
unlike the case of density perturbations the first expression for $A_{{\rm G}}$
is exact up to the slow-roll approximation.

The ratio of the gravitational to scalar
perturbations is, using Eqs.~(\ref{apprnefolds}), (\ref{contrastn})
and (\ref{gravwave}),
\be
\frac{A_{{\rm G}}}{\delta_H} =
\frac{1}{2\sqrt{\pi}} \frac{m_{{\rm Pl}}}{\alpha \Phi_0}
= \frac{1}{\sqrt{2\alpha {\cal N}}}\,.
\label{ratiogrden}
\ee
From the value of $\alpha_{50}$ from Eq.~(\ref{alfifty}), we see that the
predicted ratio of the gravitational to scalar perturbations at
$50$ $e$-foldings is
\be
\frac{A_{{\rm G}}}{\delta_H} = 6.6 \times 10^{4} \frac{m}{m_{{\rm Pl}}}\,.
\label{rationumerics}
\ee
Hence, the precise value of the mass of the zero mode determines this ratio.
Clearly the heavier field will produce more
gravitational waves relative to the scalar density contrast.
Using the bound Eq.~(\ref{massbound}), we find
\be
\frac{A_{{\rm G}}}{\delta_H} \ge 1.32 \times 10^{-5}\,.
\label{rationum}
\ee

\section{Conclusions}

We have considered dynamics of assisted chaotic inflation which can arise in
theories with large internal dimensions after compactification.  In this work,
we have focused on a simple model based on a single massive scalar field.  If
the field lives in the bulk of the fundamental theory, then upon
compactification on manifolds larger than the fundamental Planck scale, it will
give rise to a tower of massive Kaluza--Klein states.  Many of these
Kaluza--Klein
states will be lighter than the $5D$ Planck scale, and hence can be treated in
the field theory limit, where they can contribute to assisted inflation.  The
model is in agreement with COBE constraints provided that the fundamental
Planck
scale is greater than $10^{13} \, {\rm GeV}$.  The main aspects of the ensuing
assisted
dynamics are rather interesting.  Instead of the appearance of an asymptotic
attractor for the multitude of the scalar fields displaced from their
respective
minima, as in the original assisted inflation with exponential potentials
\cite{lms}, here the fields with a different mass never develop a completely
coherent motion.  Rather, the fields keep accelerating away from each other.
If
they are viewed as a collective mode, this means that there is a constant
spreading of the collective mode throughout the evolution.  However, the
spreading is very small compared to the expansion rate of the universe.
Indeed,
since the effective Hubble parameter of the universe receives contributions
from
all fields in the slow-roll regime, it is larger, and hence gives a stronger
resistance to acceleration of each field down its respective potential well.
This in turn prolongs the slow-roll regime for each field, and leads to longer
inflation overall.

Furthermore, a combination of the assisted behavior and the
higher-dimensional origin of the theory lowers very significantly
the value of the fields $\Phi$ where slow-roll ends, giving
$\Phi_{{\rm end}} \sim M^{3/2}/m^{1/2}$, rather than $m_{{\rm Pl}}$ as is usual
for inflation with fields confined to $4D$
(here $M$ is the fundamental Planck scale and $m$ the
zero mode mass). So instead of inflation terminating at the $4D$
Planck scale, it can last well below it, almost as low as the
higher-dimensional Planck scale. Therefore, to drive a long
inflation, the zero mode and all of its Kaluza--Klein siblings can
start with values of order of $m_{{\rm Pl}}$, at energy densities
far below the $4D$ Planck scale, and with values in the regime
where higher-loop supergravity corrections are much less likely to
destroy inflation. This also means that there may be
less fine-tuning in choosing the parameters of the theory. The
higher-dimensional couplings of order unity upon compactification
can naturally produce small couplings needed to satisfy the COBE
constraints, and these numbers can be perturbatively stable.

A very interesting novel feature of the assisted chaotic dynamics
is that inflationary predictions depend softly on the
initial conditions preceding the stage of inflation.
Most common models of inflation with a small number of dynamical
scalar fields exert complete amnesia on the Universe,
which forgets all about the initial state before
inflation. This is seen as a typical consequence of
cosmic no-hair theorems. However assisted chaotic inflation
appears to be more forgiving. Rather than completely washing away
all the information about the state preceding inflation, at the level of
precision we have pursued here assisted dynamics gives
a density contrast and spectral index which depend logarithmically
on the initial value of the inflationary scalars. In fact,
this effect could have been expected due to the collective nature
of the inflaton. To the subleading order in approximations, the predictions
of dynamics should recognize how many fields contributed to
inflation. On the other hand, the number of fields and the initial value of the
inflationary scalars are related. Indeed, if we start with fewer
fields higher up the potential, we may produce the same number of
$e$-foldings as if we had more fields initially closer to their
minima. Hence while both scenarios give the same picture to leading order,
they differ in the sub-leading order. Since it is at this level that the
density
perturbations are produced, clearly they will depend on the
initial values of fields and their number. The COBE normalization
permits one to eliminate the number of fields in favor of the
initial value of fields. Hence one, but not both, of these parameters can
be removed from the results for density perturbations and the
spectral index, which therefore must depend softly on the initial
value of fields. In the language of no-hair theorems, this dependence
is analogous to a kind of discrete cosmic ``hair".
Since it is soft, it will not jeopardize
the onset of inflation. However, it leads to
the possibility of getting different inflationary spectra from
theories with the same zero mode parameters in the Lagrangian,
and hence reduces the usefulness of observations
in constraining inflationary models.

It would be very interesting to study generation of density
perturbations in assisted chaotic models beyond the sub-leading
order of approximations which we pursued here. The presence of
the multitude of scalars and the absence of a late-time stable
attractor could lead to additional interesting sub-leading
corrections to the density spectrum. If such corrections are
within the observable region, they could lead to an interesting
signature of additional dimensions of the world visible
in our own sky.

\vskip1.5cm

\section*{Acknowledgments}

We thank Ed Copeland and David Wands for helpful discussions.
The work of N.K. has been supported in part by NSF Grant PHY-9870115.

\end{document}